\documentclass[final,3p,times,numbers,sort&compress]{elsarticle}
\usepackage{amsmath,amssymb,amsfonts}
\usepackage{amsthm}
\usepackage{graphicx}
\usepackage{hyperref}
\usepackage{algc}
\usepackage{xcolor}
\definecolor{mycolor}{rgb}{0.7,0.3,0.3}

\usepackage{amssymb,amsmath,graphicx}
\begin{document}

\begin{frontmatter}

%\dochead{Reply to comment}
\title{It is Useful to Analyze Correlation Graphs}

\author[LeicMath,NNU]{A.N. Gorban\corref{cor1}}
\ead{a.n.gorban@le.ac.uk}
 
\author[LeicMath,NNU]{T.A. Tyukina}
\ead{tt51@le.ac.uk}

\author[SFU]{L.I. Pokidysheva}
\ead{pok50gm@gmail.com}

\author[SFU]{E.V. Smirnova}
\ead{seleval2008@yandex.ru}

\address[LeicMath]{Department of Mathematics, University of Leicester, Leicester,   UK}
\address[NNU]{Lobachevsky University, Nizhni Novgorod, Russia}

\address[SFU]{Siberian Federal University, Krasnoyarsk, Russia}

\cortext[cor1]{Corresponding author}
\begin{abstract}
In 1987, we analyzed the changes in correlation graphs between various features of the organism during stress and adaptation. After 33 years of research of many authors, discoveries and rediscoveries, we can say with complete confidence: It is useful to analyze correlation graphs. In addition, we should add that the concept of adaptability (`adaptation energy') introduced by Selye is useful, especially if it is supplemented by `adaptation entropy' and free energy, as well as an analysis of limiting factors. Our review of these topics, ``Dynamic and Thermodynamic Adaptation Models" ({\em Phys Life Rev}, 2021, arXiv:2103.01959 [q-bio.OT]), attracted many comments from leading experts, with new ideas and new problems, from the dynamics of aging and the training of athletes to single-cell omics. Methodological backgrounds, like free energy analysis, were also discussed in depth. In this article, we provide an analytical overview of twelve  commenting papers and some related publications.
\end{abstract}  
\end{frontmatter}
\vspace{5mm}

What can be considered as the main result of a long research program? Let us take a look at great examples. According to Arnold \cite{Arnold2012}, Newton's fundamental discovery, the one which he considered necessary to keep secret and published only in the form of an anagram, in contemporary mathematical language  means: ``It is useful to solve differential equations''. 
The discovery and elaboration of a new tool, with a demonstration of its wide applicability, is a good main result when supplemented by many smaller advances that demonstrate the usefulness of the developed tool and should be important in their own right.

In 1987, we analyzed the changes in correlation graphs between various features of the organism during stress and adaptation \cite{GorSmiCorAd1st}. Results of 33 years of work of many researchers and groups were summarized in the review paper \cite{GorbanTyuPokSmi2021}. In this review, we aimed to demonstrate that  it is useful to analyze correlation graphs. In addition, we should add that the concept of adaptability (`adaptation energy') introduced by Selye in 1938 \cite{SelyeAEN} is useful, especially if it is supplemented by recently invented `adaptation entropy' and free energy, as well as by an analysis of limiting factors. 

However, the proper extended review and outlook were finalized when the twelve additional comment papers were published together with our article \cite{Franceschi2021, Bizzarri2021, Nazarenko2021, Paola2021, Vasenina2021, Zinovyev2021, Allahverdyan2021, Selvarajoo2021, Giuliani2021, LiuAihChen2021, Red2021, Ueltzh2021}.  Our `reply to comments' is an introduction to these twelve supplementary materials. There are several aspects in which commentators have enriched our review:
\begin{itemize}
\item Presentations of materials on changes in correlation networks under stress, illness and adaptation, which analyze various situations and use additional tools for data preprocessing and analysis of correlation graphs. These extensions are important for proof that analysis of correlation graphs is useful in analysis of adaptation.
\item Suggestions to extend methods onto different phenomena and problems, from aging and single cell omics to molecular networks and other intriguing systems.
\item Discussion of alternative or deeper backgrounds of the effects described in our review, and description of the conceptual problems that the proposed methodology will face in the further development and  expansion of application areas.
\end{itemize}

Of course, each comment contributed to several aspects of the content. Below, we briefly discuss these enrichments. 

\section{Additional materials on the main area of the review}

\paragraph{Dynamical network biomarkers}A treasure box of ``dynamical network biomarkers'' was opened for readers in \cite{LiuAihChen2021}. This concept was developed to characterize the critical state during the disease progression. We agree that these ideas are consistent with our correlation graph approach. Moreover, taking into account the publication dates, we are proud to mention that the dynamical network biomarkers present the further development of our approach.  Our series of publications started in 1987 (see, for example, \cite{GorSmiCorAd1st, Sedov, GorbanPokSmiTyu, GorbanSmiTyu2010, KrasnenkoPokid2010, GorbanTyukinaDeath2016}). The approach was later applied to many systems and elaborated  in various directions by many authors, with references to our work \cite{RazzhevaikinTrava1996, Giulianni2014, Bernardini2013, Censi2011, Censi2010, Razzhevaikin2012, Shchelkunova2013, SchefferEtal2012, Siggiridou2014, Cramer2016, Mojtahedi2016, Rybnikovs2017, Heiberger2018} or with rediscovery of some ideas
\cite{LonginCorrNonconst1995, Onella2, ChenAihara} (for the more detailed review we refer to \cite{GorbanTyuPokSmi2021}). Of course, each paper was not just an application or rediscovery of previously known ideas. Most groups of authors proposed their own versions of methods suitable to their material and developed some theoretical backgrounds for their findings. 

In particular, the important new tools for selection of the  relevant dominant or leading subnetworks in the large networks of attributes near crisis were developed by comparison between the internal (within the group) and external correlations.  A brief review of the theory and practice of dynamical network biomarkers was presented in the comment \cite{LiuAihChen2021}. A different approach for selection of the most relevant subnetworks of attributes is based on principal component analysis \cite{Censi2011, Censi2010}. In this approach, at the first stage, the most important principal components for disease recognition should be found (the components with the greatest significance for the classification task), and then the attributes that dominate in these principal components should be selected for further correlation and variance analysis (for more details, we refer to the original works). 

The series of work presented in \cite{LiuAihChen2021} systematically uses the concept of attractor bifurcations as a hidden cause of the observing transformation of the network. 
The general theory of bifurcation of attractors and critical delays in general dynamic systems (with and without noise) was developed in 1978-1980 in the PhD thesis of one of the authors \cite{GorbanDiss, GorbanDAN}. (An abridged version of the thesis \cite{GorbanDiss} was published later in English \cite{GorbanSingularities}, see also a brief review \cite{GorbanCommPetr2020} with further references.) However, we found that a systematic theory of the transformation of attribute networks under adaptation and stress can benefit from the concepts developed by Selye and his school. Below, we discuss this question and the non-trivial relations between attractor bifurcations and the concept adaptability  in more detail using also other comments that considered bifurcations of attractors as one of the main mechanisms of the observed effects  \cite{Bizzarri2021}. 

\paragraph{Someone's noise is another one's signal}K. Selvarajoo in his comment \cite{Selvarajoo2021} discussed high-throughput gene expressions data analyzed to interpret and predict the complex immune, cancer and biofilm responses. In general, he supported our findings but attracted more attention to biological noise. This noise can be independent from the technical noise like instrumental effects and can be a signal of approaching critical transitions. We completely agree with this comment. Moreover, the correlation graph approach discussed in our review with analysis of correlations and variances is a tool for study of sensible noise in the networks of attributes. These methods are widely used and modified for different purposes. For example, correlations and variances both in the organism-to-organism \cite{GorSmiCorAd1st, GorbanSmiTyu2010, Razzhevaikin2012}  (or cell-to-cell \cite{Selvarajoo2021}) variations and in the fluctuations of the attributes of one organism in time \cite{Cramer2016}  can provide important information about adaptation and crisis. 
The further development of this methodology provides inspiring examples even from applied humanities, such as the analysis of fluctuations in time of physiological parameters (cardiac activity and skin conductance) in a `psychotherapist-patient' pair, which measures the level of empathy and may  predict the success of psychotherapy \cite{Kleinbub2019}, or the analysis of correlations between variations of public fears for different regions of the country that can predict political crises \cite{Rybnikovs2017}. 

This topic, analysis of noise and individual behavior of various structural components (both material, like cells, tissues, etc., and virtual like principal components, biomarkers, and subnetworks) was stressed by many commentators. C. Franceschi stated that {\em ``each cell has a specific signature and present subtle differences with respect to the other cells derived from the same source,''} -- and these differences increase with age \cite{Franceschi2021}. In this comment, like in \cite{Selvarajoo2021}, our approach is friendly criticized for insufficient  attention to this heterogeneity of life system:  ``A major usually non explicit assumption of the approach adopted by Gorban et al. is that the elements of each specific cellular network are basically similar each other's. This basic assumption is likely not true.''    We fully accept the fundamental heterogeneity of many life systems at all levels but cannot accept that their homogeneity is our `major assumption'. 

However, the source of the impression that our systems are ``basically similar'' is very clear (to us): To apply the correlation graph approach to a group of biological objects (organisms, cells, etc.), no one needs their homogeneity. In fact, {\em `meta-uniformity' is desirable instead: the attributes selected for analysis must be measurable for all objects.}
Of course, some palliative solutions can be used when attributes are immeasurable for some objects, for example, a special value can be entered meaning `impossibility of measurement'. However, the correlation of the two attributes can only be calculated for groups (or subgroups) where both attributes are measurable.
Technically, one can consider a `crazy kilt' of such subgroups with a different set of objects for different pairs of attributes, but meta-homogeneity with the same attributes for all objects makes data analysis much easier and more transparent.

According to the comment \cite{Bizzarri2021}, the crucial importance of fluctuations in living systems are caused by the  ``small integer numbers of key and lock molecules'' of many types. ``The key issue is that randomness is a central feature of the microscopic level, and it is therefore unlikely that microscopic details could help in understanding the logic behind living organisms.'' Therefore, the special mesoscale approach is needed, that does not rely on the central limit theorem and mass action law but, on the other hand, is not as noisy as the microscopic level is \cite{Bizzarri2021}. Despite of that, another comment \cite{Zinovyev2021} is focused on the single cell level and discussed the data collections that are produced at that `microscopic' level by single cell omics as a promising area of applications of the correlation graph approaches and their further development.
The comment \cite{Paola2021} goes even deeper, to more `microscopic' and noisy molecular networks. 
The fluctuating world of many personalized objects can be represented as a supernetwork, that is a network of networks. Promising approaches for constructing of such supernetworks of potentially infinitely depth hierarchy were discussed in the comment \cite{Nazarenko2021}. This hierarchy starts from the graphs of binary correlations (or their close analogues) and then goes deeper to more complex structures.

Separation of real life high-dimensional and heterogeneous data into signal and noise is extremely problematic and can be never done for sure. The important data mining `axiom',  {\em Someone's noise is another one's signal}, should be included into undergraduate textbooks, and biological  networks can provide good examples of transforming apparent noise into important signal.

\section{Extending of methods to new phenomena and problems}

\paragraph{Dynamics of aging} 
According to the comment of C. Franceschi,  \cite{Franceschi2021}, the approach presented in our work ``will help not only in disentangling the complexity of aging (particularly in humans) but also in establishing a new `aging medicine', and above all in identifying new tools capable of early predicting power regarding human health trajectories.''  In this comment, the phenomena of aging that can be covered by our approach were analyzed and some open problems were formulated. 

The concept of adaptation energy and optimization of its spending  discussed in the paper ``fits very well with what I learned from the study of human aging and longevity,'' wrote one of the world's leading experts on aging. This opinion supports our claim that the concept of adaptability (`adaptation energy') is useful. It could be worth to mention that a very recent critical review of adaptation energy and its applications to exercise science
\cite{Vasenina2020} states that this concept is applicable to the phenomenon of aging rather than to sports.

C. Franceschi drafted a research program for application of dynamic and thermodynamic models of adaptation to aging phenomena. As per his comment, the main and urgent challenge to our approach, is ``the fundamental heterogeneity and uniqueness of humans at all biological level (genetic, epigenetic, metabolic, among others) which emerges whenever the results of a study are analyzed at the individual level,'' and an enormous amount of work is expecting to meet this challenge. We have already begun work on analysis large heterogeneous systems and have created open access software tools for identifying development trajectories in big data \cite{Albergante2021, ElPiGraphR, ElPiGraphMATLAB, ElPiGraphPython}. These tools can help in predicting ``human health trajectories'', as requested by the commentator. Nevertheless, there remain more inspiring questions than answers in the theory and practice of optimal aging and longevity \cite{Franceschi2021}.

\paragraph{Exercises and sport}

Exercise and sports data provide a wealth of material for the study of adaptation. Any theory of adaptation is incomplete without assimilation of these materials. We were deeply concerned that Selye's classical concept of adaptation energy did not seem to be able to describe the training phenomena that are well known in the science of exercise and beyond to anyone with a sporting background. Similar concern was formulated by several authors, see a review \cite{Vasenina2020}. Our commentators, \cite{Vasenina2021} are experts in exercise science. They analyzed the concepts of adaptation from the ``exercise science perspective''.  We are grateful for an accurate understanding of our modification of the concept of energy of adaptation and for examples that demonstrate the need for modification. Here, we would like to add one more comment. To capture the apparent training phenomena we had to introduce ``free energy of adaptation''. It included two components: adaptation energy and adaptation entropy.  
The entropy of adaptation is necessary for the formalization of adaptive changes in the scale of stressors. After training, the organism can withstand larger loads. The same load can be `small' for a trained body and `large' (or even impossible) for an untrained one. This rescaling of the stressors can be measures by a logarithm of the rescaling coefficient (logarithm is needed for additivity). This is a typical entropic measure -- logarithm of available `volume'. The dynamics of the load scale described by our free energy models \cite{GorbanTyuPokSmi2021} are qualitatively consistent with the phenomena discussed in the comment \cite{Vasenina2021}, but additional efforts are required for detailed modeling and analysis to verify the quantitative agreement. 

\paragraph{Single cell omics}

We dealt \cite{GorbanTyuPokSmi2021} with whole organisms (mice, people, plants) or ``super-organismic'' objects (stock market or social systems). A. Zinovyev proposed to change the focus  to a significantly smaller scale, to a single cell \cite{Zinovyev2021}. This is not just an interesting and important question asked by many commentators. The comment \cite{Zinovyev2021} proposed to work with single-cell omics. This is a new  technology that gives information about genomes, gene expression, and cell heterogeneity for thousands of individual cells from a single organism, simultaneously. This revolutionary technology brings a huge amount of new data we have never seen before. It is a challenge to biology and medicine to assimilate these datasets.  Several specific difficulties of single cell omics data analysis were discussed in the comment \cite{Zinovyev2021}. In particular, individual cells are usually destroyed when measured, so we can track the population of cells over time, but the state of an individual cell can only be observed once. It is very desirable 
to join individual cells (represented by points in a multidimensional space) into trajectories of their development. These trajectories can branch and the whole picture is expected to be a bunch of branching cell development and differentiation trajectories. Extraction of this bunch from a single cell data cloud is a special data mining problem and several tools are developed (for example, the method of topological grammars and principal graphs \cite{Albergante2021, ChenAlberg2019}). However, this is not the end of the story. The motion of cells in the feature space may include also oscillation in cell cycle that is different from the development and differentiation. This is especially important for growing cell populations like cancer. Special techniques are needed to separate these two types of motion: cell cycle and development with differentiation. In a recent work \cite{Aynaud2020} a motion separation approach based on independent component analysis was developed. Universal mathematical \cite{ZinovyevCellCycle2021} and computational \cite{auranicCellCycle} models of cell cycle are useful for such separation. The developed methods for processing of single cell omics data open the door for detailed analysis of adaptation and stress at the single cell level. A research program of this analysis is drafted in the comment  \cite{Zinovyev2021} and we will be happy to see the results and participate in this new fascinating intellectual travel.

\paragraph{Molecular networks}

Can we go deeper than the single cell level? The comment \cite{Paola2021} demonstrates one more step: from cells to molecular networks. Adaptation at the level of protein networks is possible.  Recent works reveal that protein networks adapt to mutations. They develop new allosteric networks.
For us, this is a completely new  angle of view, and we are more than happy that the integrated dynamic and topological approach works at that level and helps in analysis of adaptation at the level of protein networks. As a result, we can imagine the future construction of adaptation science, from the very bottom level (molecular networks) to the single cell adaptation, then the tissue and organismic levels, and finally, the level of super-organismic entities. In this context, the concept of ``network of networks'' \cite{Nazarenko2021, Whitwell2020} seems to be very useful.

\section{Alternative or deeper backgrounds of the observed effects}

\paragraph{Bifurcations versus adaptability dynamics}

In two comments, \cite{Bizzarri2021, LiuAihChen2021} the bifurcations of attractors is considered as a main mechanism of the transformation of correlation graphs near the `tipping points'. In our review  \cite{GorbanTyuPokSmi2021}, we discussed three proposed mechanisms of the observed effects: dynamics of adaptability described by the factors--resource models, approaching bifurcations of hidden dynamic systems, and approaching the border of survival (in the framework of Fokker--Planck models). All these mechanisms describe some experimental data, at least, qualitatively. Moreover, even if some data are not described, there remains a hope the the model can be modified accordingly to meet better the real phenomena. 

It is difficult to select the best type of models because all the models refer to hidden phenomena that are not observable directly. For example, the adaptation energy is not a physical quantity and cannot be measured directly. Instead, adaptation energy (or free energy) should be considered as an internal coordinate on the {\em dominant path} in the model of adaptation \cite{GorbanTyukinaDeath2016}. Many questions can be asked: What is the dynamical system that has this dominant path, what is its phase space, why the dominant path appears? (This can be the result of   either averaging, or the existence of a slow manifold, or a combination of these reasons.)

Very similar questions can be asked for the bifurcation model. Nobody has observed the underlying  dynamical systems and its `network of attractors'. Therefore, all this reasoning remains a useful metaphor. The situation is very similar to the `self-organized criticality' (SOC) that provides us by a metaphor and  simple nice examples like a sandpile but does not create quantitative models. 
``One of the main misconceptions about SOC is to think that it is a traditional quantitative theory and that, as such, it can yield detailed quantitative predictions once confirmed that the system exhibits SOC dynamics \cite{Sanchez2015}.'' 

Existence of dominant path and a single parameter that describes behavior of a complex system is usual near bifurcation and phase transitions. Perhaps, the first famous example gave Lev Landau in his work about stability loss of laminar flows \cite{Landau1944}. In this work, he found the equation for the deviations from the steady laminar flow without explicit use of the Navier-Stokes equations. The closeness to the bifurcation was the main assumption. In that sense, two approaches, bifurcation and a special dominant path with adaptability as a main coordinate, are not in principal contradictions. Fluctuations and critical delays become very important near bifurcations, here we agree with our commentators. (See also our detailed analysis of attractor bifurcations and critical delays in dynamical systems, \cite{GorbanDiss, GorbanDAN, GorbanSingularities, GorbanCommPetr2020}. In particular, we analyzed effect of fluctuations modeled by $\varepsilon$-trajectories.)

Despite of some similarity, there exists a difference between just a general `closeness to a bifurcation' and a specific dominant path factor-resource models. Instead of rephrasing of our large paper, we will present a very recent simple example of a social system, where these factor-resource relations are obvious. 

Development of world-class universities in China was analyzed in \cite{HartleyUniver2021}. This process was started in 1998 and recently, high positions of Chinese universities in world ranking are widely known. At the beginning of this race for ranking, vast investments of money have indeed earned China’s universities `quick wins'. In that sense, one factor dominated the accelerated development of the universities to the relatively high level. But after that, more individual institutional and cultural factors determine the development of the universities, ``including dominion over university-level strategic initiatives, promotion, tenure, research directions, establishment of new academic programs, and publication and academic content'' \cite{HartleyUniver2021}. Thus, efforts and successes of different universities become different and the race for higher success is not dominated just by funding. According to  \cite{HartleyUniver2021}, the simple government's command-and-control approach to higher education institutes became largely inconsistent  with the diverse world of universities and at this ``multifactorial'' stage, a more flexible approach is necessary. We cannot comment this advices abut management, but the difference between the single-factor development and polyphonic multifactorial motion seems to be obvious.

The comments \cite{Bizzarri2021, LiuAihChen2021} include much more material. In particular,  \cite{Bizzarri2021} presents a very dense review of many mathematical, biological and even epistemological  problems in understanding biological adaptation with special attention to stochastic processes and limited predictive power of many classical families of models like ordinary differential equations of chemical kinetics. (The comment \cite{Selvarajoo2021} also stressed that these models are insufficient in the stochastic world of genome activity.) High importance of fluctuations was  demonstrated also in the comment \cite{LiuAihChen2021}. All these problems are very important, indeed. Here we would like just to stress that relatively simple `middle level' models may work well despite the complexity and stochasticity of the low-level detailed picture. 

\paragraph{Statistical physics of life and network thermodynamics}

The factor-recourse models work like thermodynamics, whereas the hidden bifurcation of unknown attractors is a metaphor of statistical physics. A. Giuliani in his comment  \cite{Giuliani2021} presented an idea of `organized complexity' and `statistical physics of life'. He proposed that the foundation of a biological statistical mechanics ``should focus on the changes in correlation patterns marking state transitions''.  At the same time, the network thermodynamics of life is necessary to consider changes in macroparameters. For this purpose, we utilized the typology of systems of factors and their interactions proposed by Tilman \cite{Tilman1980}. We  developed these ideas further and applied to analysis of adaptation. In addition to the Liebig's law of the minimum, various different interaction of factors are possible like synergy. The rich family of ``factor-resource'' models can work not only as a metaphor, but also provide quantitative models of real systems.
To overcome the quasistatic nature of these models (`periodic adaptation'), a family of dynamic models of individual adaptation was proposed. 

The manifesto of statistical physics of life \cite{Giuliani2021} supports the main results: 
It is useful to analyze correlation graphs and thermodynamic models based on adaptability (`adaptation energy') concept. It also opens further methodological perspectives.

 \paragraph{Networks of networks}

The concept of ``networks of networks'' covers many ideas expressed by the commentators. The hierarchy  with  molecular networks \cite{Paola2021} at the low level, then networks of cells and their features \cite{Zinovyev2021}, then correlation networks of the attributes of an organisms, and so on, presents an example of such a network of networks. Analyzing these constructs may require special mathematical and statistical tools.

In our paper \cite{GorbanTyuPokSmi2021}, we considered the correlation graphs, where the nodes were attributes and edges represented the correlations between the attributes. The simplest definition of edge weight is Pearson's correlation coefficient. In our previous works, we tried many various characteristics and, finally, were satisfied by the simplest version. Nevertheless, this is not a compulsory choice and for formal description of `networks of networks' various different approaches may be proposed. Informational characteristics, like information gain, may be useful, especially for qualitative attributes. Various regression residuals are used in the ``parenclitic'' (from the Greek for ``deviation'') and synolytic (from the Greek  for ``ensemble'')   network representations \cite{Nazarenko2021}. These networks were successfully used in medical applications.  Essentially, they are also the correlation networks but with the different measures of correlations. However, the details matter and different measures may be useful in different situations. It will be important to compare different approaches and we hope that  the comment \cite{Nazarenko2021} could be extended to an analytical work about different versions of correlation graphs and their performance analysis. Even more, the ensemble of different correlation graphs can be used for description of a complex biological system, and we may not need to choose the best correlation graph, but rather use all of them together  \cite{Nazarenko2021}.

Another technique is discussed in the comment \cite{Red2021}. Many modern tools are developed for learning of intellectual agents, including neural networks. Adaptation can be modeled by learning. Evolution of a population of such agents with small mutations of `genotypes' and  optimization of phenotypes by learning can clarify the problem of instantaneous fitness discussed in our work. These systems of intellectual agents can be used for modeling networks of networks.
 
\paragraph{Three (or four?) faces of free energy in adaptation and stress}

Two free energies were used in our work. One of them is the classical physico-chemical free energy used in \cite[Appendix A]{GorbanTyuPokSmi2021} for demonstration of `thermodynamic cost of homeostasis'. Another is the adaptation free energy, combined from the adaptation energy and the adaptation entropy introduced for modeling of the adaptive rescaling of stressors in training. Two comments, \cite{Allahverdyan2021, Ueltzh2021} added a lot to this topic.

The main challenge, formulated in \cite{Allahverdyan2021} is ``to establish a real correspondence between axiomatic structures of adaptation and microscopic models.''
According to this comment, the difference between homeostasis and adaptation is the difference between two types of control: for homeostasis, this is control of dynamic variables (`thermodynamic coordinates'), like concentrations, and for adaptation this is control of the conjugated variables (`thermodynamic forces'), like chemical potentials. Behind this definition, we found the representation of biological systems as {\em controlled dynamical system of non-equilibrium thermodynamics}. We fully support this basic idea. Adaptation in dynamical systems \cite{Tyukin2011} is important for analysis of biological processes, and these dynamical  systems must obey thermodynamics.  However, realization of this idea still requires a lot of work and further details, even in the basic definitions.

The comment  \cite{Allahverdyan2021} presents  Bauer's axiomatics of adaptation \cite{Bauer1935} and explains, why it meets the physico-chemical backgrounds of biological adaptation better  than Selye  (or Selye--Goldstone) axioms. 
A research program ``towards the physics of adaptation'' is proposed that consists of two items. We reformulate them as follows:
\begin{enumerate}
\item Develop accurate thermodynamic estimates of energy consumption in adaptation and homeostasis (according to \cite{Allahverdyan2021}, the existing models overestimate this consumption, especially for adapted steady states). 
\item Integrate the Selye--Goldstone phenomenology with the physico-chemical approach to adaptation proposed by Bauer (and the modern versions of these approaches) and create more realistic and interesting models.
\end{enumerate}
 
It seems that the comment \cite{Ueltzh2021} approaches thermodynamic of adaptation from the opposite side. The concept of variational free energy is clearly explained from scratch, from variational Bayesian inference. The formal definition \cite[Box 1]{Ueltzh2021} is followed by an explanation of qualitative details, meaningful analogies and even metaphors. According to \cite{Ueltzh2021}, the variational free energy evaluate the surprisal under an optimism-biased expectations from the world,  that is, it measures instantaneous distress. It can be used for deriving the instantaneous fitness using the learning metaphor of adaptation (compare to the remark from \cite{Red2021} about evolutionary learning of intellectual agents and models of adaptation and evolution). It is impressive that the dynamics of ``Bayesian thermostats'' in active inference demonstrate the behavior of variance and correlations between intellectual agents that was observed in many living systems under stress in adaptation  process.

Finally, the concept of variational free energy meets physical thermodynamics in the following form: in practice of modeling living systems, the variational free energy and the storaged of thermodynamic free energy have {\em inverse relationships} because ``the minimisation of variational free energy in living agents is associated with the build-up and maintenance of the physical structure realizing these agents.''

The combination of comments \cite{Allahverdyan2021, Ueltzh2021} gives hope that a new revolutionary concept of variational thermodynamics of adaptation will soon appear, taking into account both non-equilibrium thermodynamics and Bayesian learning.

A very minor answer to a minor comment from   \cite{Ueltzh2021} is ethically important for us. In a brief discussion of our estimate of free energy cost of a non-equilibrium state \cite[Appendix A]{GorbanTyuPokSmi2021}, an earlier work of Horowitz et al. (2017) \cite{Horowitz2017} was mentioned in \cite{Ueltzh2021}. We were ready to cite our book (1984), which contains these estimates (and much more) \cite{GorbanObkhod}
(some results of this book were later published in English  \cite{GorbanKaganovich2006}), but we were ashamed when found that  the first results of this type were published in 1935 by Bauer \cite{Bauer1935} (not cited in our review). We must never forget this very talented researcher who was arrested and killed on the peak of his creative life. 

\section{Instead of conclusion} The combination of comments creates a polyphonic effect around a very deep problem of adaptation. In this `reply' we tried to share our impressions and enjoyment of this polyphony. We are very grateful to all commentators for new questions, tools, hypotheses, and answers to some problems. The main expected result of our joint efforts will be ``the enlargement of the spectrum of the phenomena amenable of a quantitative approach'' \cite{Giuliani2021}. 

\section*{Acknowledgments}
The work was supported by the Ministry of Science and Higher Education of the Russian Federation (Project No. 075-15-2021-634).
.


\section*{References}

 \begin{thebibliography}{99}

\bibitem{Arnold2012}Arnold VI. {\em Geometrical methods in the theory of ordinary differential equations.} New York: Springer,   2012.

\bibitem{GorSmiCorAd1st}Gorban AN,  Manchuk VT, 
Petushkova (Smirnova) EV. Dynamics of physiological paramethers
correlations and the ecological-evolutionary principle of
polyfactoriality, In: Problemy Ekologicheskogo Monitoringa i
Modelirovaniya Ekosistem [The Problems of Ecological Monitoring and
Ecosystem Modelling], Vol. 10. Gidrometeoizdat, Leningrad (1987), p. 187--198.


\bibitem{GorbanTyuPokSmi2021}Gorban AN, Tyukina TA, Pokidysheva LI, Smirnova EV. Dynamic and thermodynamic models of adaptation. {\em Phys Life Rev} 2021;37:17--64. \url{https://doi.org/10.1016/j.plrev.2021.03.001}. 

\bibitem{SelyeAEN}Selye H. Adaptation Energy. Nature 1938;141(3577):926.
\url{https://doi.org/10.1038/141926a0}


\bibitem{Franceschi2021}Franceschi C.
Aging, Inflammaging and Adaptation: Comment on ``Dynamic and thermodynamic models of adaptation'' by A.N. Gorban et al.
{\em Phys  Life Rev} 2021;38:107--110. \url{https://doi.org/10.1016/j.plrev.2021.07.001}

\bibitem{Bizzarri2021}Bizzarri M, Pontecorvi P.
Critical transition across the Waddington landscape as an interpretative model: Comment on ``Dynamic and thermodynamic models of adaptation'' by A.N. Gorban et al.
{\em Phys  Life Rev} 2021;38:115--119. \url{https://doi.org/10.1016/j.plrev.2021.05.010}

\bibitem{Nazarenko2021}Nazarenko T, Blyuss O, Whitwell H, Zaikin A.
Ensemble of correlation, parenclitic and synolitic graphs as a tool to detect universal changes in complex biological systems: Comment on ``Dynamic and thermodynamic models of adaptation'' by A.N. Gorban et al.
{\em Phys  Life Rev} 2021;38:120--123. \url{https://doi.org/10.1016/j.plrev.2021.05.009}

\bibitem{Paola2021}Di Paola L, Leitner DM.
Network models of biological adaptation at the molecular scale: Comment on ``Dynamic and thermodynamic models of adaptation'' by A.N. Gorban et al. {\em Phys  Life Rev} 2021;38:124--126. \url{https://doi.org/10.1016/j.plrev.2021.05.008}

\bibitem{Vasenina2021}Vasenina E, Kataoka R, Loenneke JP, Buckner SL.
Exercise science perspective: Comment on ``Dynamic and thermodynamic models of adaptation'' by Alexander N. Gorban et al.
{\em Phys  Life Rev} 2021;38:129--131. \url{https://doi.org/10.1016/j.plrev.2021.05.005}

\bibitem{Zinovyev2021}Zinovyev A.
Adaptation through the lens of single-cell multi-omics data: Comment on ``Dynamic and thermodynamic models of adaptation'' by A.N. Gorban et al.
{\em Phys  Life Rev} 2021;38:132--134.
\url{https://doi.org/10.1016/j.plrev.2021.05.004}

\bibitem{Allahverdyan2021}Allahverdyan AE.
Energy dissipation and storage in adaptation and homeostasis: Comment on ``Dynamic and thermodynamic models of adaptation'' by Alexander N. Gorban et al.
{\em Phys  Life Rev} 2021;38:137--139.
\url{https://doi.org/10.1016/j.plrev.2021.05.002}

\bibitem{Selvarajoo2021}Selvarajoo K.
Searching for unifying laws of general adaptation syndrome: Comment on ``Dynamic and thermodynamic models of adaptation'' by Alexander N. Gorban et al.
{\em Phys  Life Rev} 2021;38:97--99.
\url{https://doi.org/10.1016/j.plrev.2021.04.001}

\bibitem{Giuliani2021}Giuliani A.
The statistical mechanics of life: Comment on ``Dynamic and thermodynamic models of adaptation'' by Alexander N. Gorban et al.
{\em Phys  Life Rev} 2021;38:100--102.
\url{https://doi.org/10.1016/j.plrev.2021.04.003}

\bibitem{LiuAihChen2021}Liu R, Aihara K, Chen L.
Collective fluctuation implies imminent state transition: Comment on “Dynamic and thermodynamic models of adaptation'' by Alexander N. Gorban et al.
{\em Phys  Life Rev} 2021;38:103--107.
\url{https://doi.org/10.1016/j.plrev.2021.04.002}

\bibitem{Red2021}Red'ko VG.
Some aspects of adaptation and evolution: Comment on ``Dynamic and thermodynamic models of adaptation'' by Alexander N. Gorban et al.
{\em Phys  Life Rev} 2021;38:108--110.
\url{https://doi.org/10.1016/j.plrev.2021.04.004}

\bibitem{Ueltzh2021}Ueltzhöffer K., Da Costa L, Friston KJ.
Variational free energy, individual fitness, and population dynamics under acute stress: Comment on “Dynamic and thermodynamic models of adaptation'' by Alexander N. Gorban et al.
{\em Phys  Life Rev} 2021;38:111--115.
\url{https://doi.org/10.1016/j.plrev.2021.04.005}

\bibitem{Sedov}Sedov KR,  Gorban' AN,  Petushkova (Smirnova) EV,  Manchuk VT, 
Shalamova EN.  Correlation adaptometry as a method of screening
of the population. Vestn Akad Med Nauk SSSR 1988;(10):69--75. PMID:
3223045 

 \bibitem{GorbanPokSmiTyu}Gorban AN,  Pokidysheva LI, 
    Smirnova EV,  Tyukina TA. Law of the minimum paradoxes, Bull Math Biol 2011;73(9):2013--2044. \url{https://doi.org/10.1007/s11538-010-9597-1}
    
\bibitem{GorbanSmiTyu2010}Gorban AN, Smirnova EV, Tyukina TA.   Correlations, risk and crisis: from physiology to finance.
    Physica A,  2010;389(16):3193--3217.  \url{https://doi.org/10.1016/j.physa.2010.03.035}
  

\bibitem{KrasnenkoPokid2010}Krasnenko, A.N., Pokidysheva, E.V., Veretnova, K.Y., Tyukina, T.A.: Analysis of Correlations in the Russian Banking System in Adapting to the Economic Crisis of 2007-2008. Journal of Siberian Federal University. Mathematics and Physics 3(4), 521–532 (2010) 

\bibitem{Pokidysheva2013}Pokidysheva L, Ignatova I. Principal component analysis used in estimation of human's immune system, suffered from allergic rhinosinusopathy complicated with clamidiosis or without it.  In: Kountchev R., Iantovics B. (eds) Advances in Intelligent Analysis of Medical Data and Decision Support Systems. Series Studies in Computational Intelligence, vol 473. Springer, Heidelberg, 2013, 147--156. \url{https://doi.org/10.1007/978-3-319-00029-9_13}

\bibitem{GorbanTyukinaDeath2016}Gorban AN, Tyukina TA, Smirnova EV,  Pokidysheva LI.  Evolution of adaptation mechanisms: Adaptation energy, stress, and oscillating death. J Theor Biol, 2016;405:127--139. \url{https://doi.org/10.1016/j.jtbi.2015.12.017}
     
\bibitem{RazzhevaikinTrava1996}Karmanova IV,  Razzhevaikin VN, 
Shpitonkov MI.  Application of correlation adaptometry for
estimating a response of herbaceous species to stress loadings,
Doklady Botanical Sciences, 1996;346--348:4--7. [Translated from
Doklady Akademii Nauk SSSR, 346, 1996.]

\bibitem{Giulianni2014}Giuliani A.  Statistical mechanics of gene expression networks: Increasing connectivity as a response to stressful condition. Adv. Syst. Biol., 2014;3:1-4.

\bibitem{Bernardini2013}Bernardini C, Censi F, Lattanzi W, Barba M, Calcagnini G, Giuliani A,   Tasca G,  Sabatelli M, Ricci E,  Michetti F. Mitochondrial Network Genes in the Skeletal Muscle of Amyotrophic Lateral Sclerosis Patients. PLoS ONE 2013;8(2):e57739. \url{https://doi.org/10.1371/journal.pone.0057739}


\bibitem{Censi2011}Censi F, Giuliani A, Bartolini P,  Calcagnini G. A multiscale graph theoretical approach to gene regulation networks: A case study in atrial fibrillation. IEEE Trans. Biomed. Eng.  2011;58(10)2943--2946, 2010.  \url{https://doi.org/10.1109/TBME.2011.2150747}

  \bibitem{Censi2010}Censi F, Calcagnini G, Bartolini P, Giuliani A. A Systems Biology Strategy on Differential Gene Expression Data Discloses Some Biological Features of Atrial Fibrillation. PLoS ONE, 2010; 5(10): e13668. \url{https://doi.org/10.1371/journal.pone.0013668}.


 \bibitem{SchefferEtal2012}Scheffer M, Carpenter SR, Lenton TM, Bascompte J,  Brock W,  Dakos V, van de Koppel J, van de Leemput IA,  Levin SA,
    van Nes EH, Pascual M,  Vandermeer J. Anticipating critical transitions.
    Science,  2012;338(6105):344--348. \url{https://doi.org/10.1126/science.1225244}
    
\bibitem{Razzhevaikin2012}Razzhevaikin VN, Shpitonkov MI. The model of correlation adaptometry and its use for estimation of obesity treatment efficiency. Russian Journal of Numerical Analysis and Mathematical Modelling. 2012 Apr 1;26(6):565-74.

\bibitem{Shchelkunova2013}Shchelkunova TA, Morozov IA, Rubtsov PM, Bobryshev YV, Sobenin IA, Orekhov AN, et al.  Lipid Regulators during Atherogenesis: Expression of LXR, PPAR, and SREBP mRNA in the Human Aorta. PLoS ONE 2013;8(5): e63374. \url{https://doi.org/10.1371/journal.pone.0063374}

\bibitem{Siggiridou2014}Siggiridou E, Kugiumtzis D, Kimiskidis VK. Correlation networks for identifying changes in brain connectivity during epileptiform discharges and transcranial magnetic stimulation. Sensors 2014;14(7):12585-97. \url{https://doi.org/10.3390/s140712585}

\bibitem{Cramer2016} Cramer AOJ,  van Borkulo CD ,   Giltay EJ,   van der Maas HLJ,   Kendler KS,   Scheffer M,  Borsboom D, Major Depression as a Complex Dynamic System. PloS One 2016;11(12):e0167490. \url{https://doi.org/10.1371/journal.pone.0167490}

\bibitem{Mojtahedi2016}Mojtahedi M, Skupin A, Zhou J, Castano IG, Leong-Quong RYY, Chang H, Trachana K, Giuliani A, Huang S.  Cell Fate Decision as High-Dimensional Critical State Transition. PLoS Biol 2016;14(12):e2000640. \url{https://doi.org/10.1371/journal.pbio.2000640}  

\bibitem{Rybnikovs2017}Rybnikov SR, Rybnikova NA,  Portnov BA. Public Fears in Ukrainian Society: Are Crises Predictable? Psychol Dev Soc 2017;29(1):98--123.
\url{https://doi.org/10.1177/0971333616689398}

\bibitem{Heiberger2018}Heiberger RH. Predicting economic growth with stock networks. Physica A 2018;489:102-11. \url{https://doi.org/10.1016/j.physa.2017.07.022}

\bibitem{LonginCorrNonconst1995}Longin F,  Solnik B, Is the correlation in
international equity returns constant: 1960-1990? J. Int Money Finance 1995;14(1):3--26.
\url{https://doi.org/10.1016/0261-5606(94)00001-H}

\bibitem{Onella2}Onnela J-P, Kaski K,  Kert\'esz J.  Clustering and information in correlation based financial networks. Eur. Phys. J. B 2004;38:353--362.
\url{https://doi.org/10.1140/epjb/e2004-00128-7} 

\bibitem{ChenAihara}Chen L, Liu R, Liu ZP, Li M, Aihara K. Detecting early-warning signals for sudden deterioration of complex diseases by dynamical network biomarkers. Sci Rep 2012;2:1--8.
\url{https://doi.org/10.1038/srep00342}


\bibitem{GorbanDiss}Gorban A.N.,  Slow relaxations and bifurcations of omega-limit sets
of dynamical systems, PhD Thesis in Physics \& Math (Differential Equations \& Math.Phys), 
Kuibyshev, Russia, 1980.

\bibitem{GorbanDAN}Gorban AN,  Cheresiz VM. Slow relaxations of dynamical systems and bifurcations of $\omega$-limit sets. Dokl. Akad. Nauk SSSR 1981;261(5)1050--1053, communicated by S.L. Sobolev [English Translation: Sov. Math., Dokl. 1981 24:645--649].  

\bibitem{GorbanSingularities}Gorban AN. Singularities of transition processes in dynamical systems: Qualitative theory of critical delays. Electr. J. Diff. Eqns., Monograph 05, 2004. \url{https://ejde.math.txstate.edu/Monographs/05/gorban.pdf}

\bibitem{GorbanCommPetr2020}Gorban AN. Singularities of transient processes in dynamic and beyond. Comment  on ``Long transients in ecology: Theory and applications'' by Sergei Petrovskii et al. {\em Phys Life Rev}  2020;32:46--49. \url{https://doi.org/10.1016/j.plrev.2019.12.002}

\bibitem{Kleinbub2019}Kleinbub JR, Palmieri A, Orsucci FF, Andreassi S, Musmeci N, Benelli E, Giuliani A, de Felice G. Measuring empathy: A statistical physics grounded approach. Physica A 2019;526:120979. \url{https://doi.org/10.1016/j.physa.2019.04.215}

\bibitem{Vasenina2020}Vasenina E, Kataoka R, Buckner SL. Adaptation energy: experimental evidence and applications in exercise science. J Trainology 2020;9:66--70. \url{https://doi.org/10.17338/trainology.9.2_66}

\bibitem{Albergante2021}Albergante L, Mirkes E, Bac J, Chen H, Martin A, Faure L, Barillot E, Pinello L, Gorban A, Zinovyev A. Robust and Scalable Learning of Complex Intrinsic Dataset Geometry via ElPiGraph. Entropy  2020;22(3):296. \url{https://doi.org/10.3390/e22030296}
\bibitem{ElPiGraphR}ElPiGraph in R on github, 2018. \url{https://github.com/sysbio-curie/ElPiGraph.R}
\bibitem{ElPiGraphMATLAB}ElPiGraph in MATLAB on github, 2020. \url{https://github.com/sysbio-curie/ElPiGraph.M}
\bibitem{ElPiGraphPython}ElPiGraph in Python on github, 2020. \url{https://github.com/sysbio-curie/ElPiGraph.P}


\bibitem{ChenAlberg2019}Chen H, Albergante L, Hsu JY, Lareau CA, Lo Bosco G, Guan J, et al. Single-cell trajectories reconstruction exploration and mapping of omics data with STREAM. Nat Commun 2019;10(1):1903. \url{https://doi.org/10.1038/s41467-019-09670-4}.


\bibitem{Aynaud2020}Aynaud MM, Mirabeau O, Gruel N, Grossetete S, Boeva V, Durand S, et al. Transcriptional programs define intratumoral heterogeneity of Ewing sarcoma at single-cell resolution. Cell Rep 2020;30(e6):1767–79. \url{https://doi.org/10.1016/j.celrep.2020.01.049}.

\bibitem{ZinovyevCellCycle2021}Zinovyev A, Sadovsky M, Calzone L, Fouch\'{e} A, Groeneveld CS, Chervov A, Barillot E, Gorban AN. Modeling progression of single cell populations through the cell cycle as a sequence of switches. bioRxiv  2021;2021.06.14.448414. \url{https://doi.org/10.1101/2021.06.14.448414}

\bibitem{auranicCellCycle}Zinovyev A, Sadovsky M, Calzone L, Fouch\'{e} A, Groeneveld CS, Chervov A, Barillot E, Gorban AN. Python notebooks for the article ``Modeling progression of single cell populations through the cell cycle as a sequence of switches.''  github 2021. \url{https://github.com/auranic/CellCycleTrajectory_SegmentModel}

\bibitem{Whitwell2020}Whitwell HJ, Bacalini MG, Blyuss O, Chen S, Garagnani P, Gordleeva SY, et al. The human body as a super network: digital methods to analyze the propagation of aging. Front Aging Neurosci 2020;12:136. \url{https://doi.org/10.1016/j.plrev.2021.03.001}


\bibitem{Sanchez2015}Sanchez R, Newman DE. Self-organized criticality and the dynamics of near-marginal turbulent transport in magnetically confined fusion plasmas. Plasma Physics and Controlled Fusion  2015;57(12):123002. \url{https://doi.org/10.1088/0741-3335/57/12/123002}

\bibitem{Landau1944}Landau LD. On the problem of turbulence. Dokl  Akad  Nauk SSSR, 1944;44:339--342. [Engl. transl. in C. R. Acad. Sci. URSS,  1944;44:311--314, and in Collected Papers. Pergamon Press, Oxford and Gordon and Breach, New York (1965), pp. 387--391.]
\url{https://doi.org/10.1016/b978-0-08-010586-4.50057-2}


\bibitem{HartleyUniver2021}Hartley K, Jarvis DS. Let nine universities blossom: opportunities and constraints on the development of higher education in China. High  Educ  Res Dev  2021:1--5. \url{https://doi.org/10.1080/07294360.2021.1915963}


\bibitem{Tilman1980}Tilman D.  Resources: a graphical-mechanistic approach to
competition and predation. Am. Nat. 1980;116(3);362--393.
\url{https://doi.org/10.1086/283633}

\bibitem{Tyukin2011}Tyukin I. Adaptation in dynamical systems. NY, Cambridge University Press (2011).

\bibitem{Bauer1935}Bauer ES. Theoretical biology. Moscow, VIEM  (1935). [Reprinted: Budapest, Akad\'{e}miai Kiad\'{o} (1984)]

\bibitem{Horowitz2017}Horowitz JM, Zhou K, England JL. Minimum energetic cost to maintain a target nonequilibrium state. Phys Rev E 2017;95(4):042102. \url{https://doi.org/10.1103/PhysRevE.95.042102}

\bibitem{GorbanObkhod}Gorban AN. Equilibrium Encircling. Equations of Chemical Kinetics and their Thermodynamic Analysis, Novosibirsk, Nauka Publ. (1984).

\bibitem{GorbanKaganovich2006}Gorban AN, Kaganovich BM, Filippov SP, Keiko AV,  Shamansky VA, Shirkalin IA. Thermodynamic Equilibria and Extrema: Analysis of Attainability Regions and Partial Equilibria. Springer, Berlin-Heidelberg-New York (2006).



 \end{thebibliography}
 \end{document}